\documentclass[twocolumn,dvipsnames]{aastex62}
\pdfoutput=1 
\usepackage{amsmath,amstext}
\usepackage{apjfonts}
\usepackage[figure,figure*]{hypcap}
\usepackage{ulem}
\usepackage{xspace}

\usepackage{graphicx}
\usepackage{natbib}
\usepackage{amssymb}
\usepackage{amsmath}
\usepackage{array,booktabs}
\usepackage{mathptmx}
\usepackage{booktabs}
\usepackage{hyperref}
\usepackage{float}

\newcommand{\msun}{\,{\rm M_{\odot}}}
\newcommand{\cm}{\,{\rm cm}}
\newcommand{\s}{\,{\rm s}}	
	
\newcommand{\erg}{\,{\rm erg}}

\newcommand{\mbh}{\,{M_{\rm BH}}}

\DeclareSymbolFont{cmletters}{OML}{cmm}{m}{it}
\DeclareMathSymbol{v}{\mathalpha}{cmletters}{"76}

\usepackage{xcolor}

\shorttitle{Fast Transients from Magnetic Disks Around Non-Spinning Collapsar Black Holes}
\shortauthors{Bopp \& Gottlieb}

\begin{document}
 \title{Fast Transients from Magnetic Disks Around Non-Spinning Collapsar Black Holes}

    \author[0009-0006-1778-5715]{Justin Bopp}
	\email{jbopp@gradcenter.cuny.edu}
    \affiliation{Department of Astrophysics, the Graduate Center, City University of New York, 365 Fifth Ave., New York, NY 10016, USA}

    \author[0000-0003-3115-2456]{Ore Gottlieb}
	\email{ogottlieb@flatironinstitute.org}
    \affiliation{Center for Computational Astrophysics, Flatiron Institute, 162 5th Avenue, New York, NY 10010, USA}
    \affil{Department of Physics and Columbia Astrophysics Laboratory, Columbia University, Pupin Hall, New York, NY 10027, USA}

\begin{abstract}

Most black holes (BHs) formed in collapsing stars have low spin, though some are expected to acquire a magnetic accretion disk during the collapse. While such BH disks can launch magnetically driven winds, their physics and observational signatures have remained unexplored. We present global 3D general relativistic magnetohydrodynamic simulations of collapsing stars that form slowly spinning BHs with accretion disks. As the disk transitions to a magnetically arrested state, it drives mildly relativistic, wobbling, collimated magnetic outflows through two mechanisms: steady outflows along vertical magnetic field lines (``Blandford-Payne jets'') and magnetic flux eruptions. With isotropic-equivalent energy of $E_{\rm iso}\approx10^{52}\,{\rm erg}$, exceeding that of relativistic jets from BHs with spin $a\lesssim 0.25$, the disk outflows unbind the star, ultimately capping the final BH mass at $ M_{\rm BH} \approx 4\,M_\odot$. Once the outflows emerge from the star, they produce mildly relativistic shock breakout and cooling emission. Our cooling emission estimates suggest a bright near-ultraviolet and optical signal at absolute magnitude $M_{\rm AB}\approx-16$ lasting for several days. This indicates that disk winds could be responsible for the first peak in the double-peaked light curves observed in Type Ib/c supernovae (SNe) or power another class of transients. The detection rate in the upcoming Rubin Observatory and ULTRASAT/UVEX will enable us to differentiate between competing models for the origin of the first SN peak and provide constraints on the physics and formation rate of accretion disks in core-collapse SNe.

\end{abstract}

\section{Introduction}

Collapsing massive stars are the primary channel for black hole (BH) formation in the Universe \citep[e.g.,][]{Heger2003,Smartt2015,Kochanek2015}. When a massive star’s core exhausts its nuclear fuel, it collapses under its own gravity, forming a protoneutron star (PNS) that may eventually collapse into a BH \citep[e.g.,][]{OConnor2011,Sukhbold2016,Obergaulinger2020}. These newly formed BHs can power diverse outflows, each with unique electromagnetic signatures that offer insights into their physical properties.

One of the most well-known transients powered by BHs in collapsing stars is long-duration $\gamma$-ray bursts \citep[lGRBs;][]{Woosley1993,MacFadyen&Woosley1999,MacFadyen2001}, generated by electromagnetically driven jets from magnetized BHs via the ``Blandford-Znajek'' mechanism \citep[hereafter BZ;][]{Blandford&Znajek1977}. While all supernovae (SNe) coinciding with lGRBs are of Type Ic-BL \citep[e.g.,][]{Cano2017}, there is a significant discrepancy between the observed rates of lGRBs and Type Ic-BL SNe. The beaming-corrected lGRB rate is $ {\mathcal{R}_{\rm lGRB}}\sim 100 ~{\rm Gpc^{-3}}~{\rm yr}^{-1} $ \citep{Wanderman2010}, whereas the Type Ic-BL SNe rate is $ {\mathcal{R}_{\rm Ic-BL}}\sim 5000 ~{\rm Gpc^{-3}}~{\rm yr}^{-1} $ \citep{Perley2020,Frohmaier2021}. This disparity implies that most Type Ic-BL SNe do not produce jets that emerge from the collapsing star. The absence of such jets can be explained by a combination of two physical processes:

(i) Jets fail to escape: Jets may form in most, if not all, Type Ic-BL progenitors, but many fail to break through the stellar envelope, preventing a lGRB signal from being generated \citep{Mazzali&Valenti&DellaValle2008,Bromberg&Nakar&Piran2011,Bromberg2012}. Even when jets are choked, they still energize a hot, mildly relativistic cocoon, potentially unbinding the stellar envelope and producing fast transients \citep[e.g.,][]{Ramirez-Ruiz2002,Margutti2014,Nakar2015,Sobacchi2017,Irwin2019,DeColle2022,Eisenberg2022}. These events may be linked to phenomena such as:
Low luminosity GRBs (llGRBs), whose rate might be considerably higher than that of lGRBs \citep{Coward2005, Cobb2006, Pian2006, Soderberg2006a, Liang2007, Guetta&DellaValle2007, Fan2011}; and
fast blue optical transients \citep[FBOTs;][]{Drout2014, Margutti2019}, which have been demonstrated to be consistent with cocoon emission \citep{Gottlieb2022b}.

(ii) Jets fail to launch: Jets can only form in progenitor stars that produce moderately spinning BHs threaded by strong magnetic fields and surrounded by accretion disks \citep[e.g.,][]{Symbalisty1984,Woosley1993, Metzger2008, Gottlieb2022a}. If the core spins rapidly enough to form a millisecond PNS with an accretion disk, the newly formed PNS will generate strong magnetic fields, which will be anchored to the BH horizon by the disk upon PNS's collapse, enabling BZ jet launching \citep{Gottlieb2024b}. However, efficient angular momentum transport within stars often leads to slowly spinning cores, resulting in natal BHs with low spin parameters, $a \approx 10^{-2}$ \citep{Fuller2019, Belczynski2020}. Even the fastest-spinning BHs capable of powering lGRBs tend to have moderate spin values of $0.2 \lesssim a \lesssim 0.5$ \citep{Gottlieb2023}, suggesting that most BHs possess spins of $ a \ll 1 $. 

Nevertheless, slowly spinning BHs are expected, at least in some cases, to form with an accompanying accretion disk, which could drive outflows that may even explode the star \citep{Hayakawa2018,Quataert2019}. For example, massive compact progenitors possess a larger reservoir of angular momentum at greater radii \citep[e.g.,][]{Woosley&Heger2006,Woosley2012}, implying that the BH is initially born slowly spinning, and may later form an accretion disk. In fact, even a random velocity field in the stellar convective region might exhibit some degree of coherence, potentially contributing to late disk formation \citep{Quataert2019,Antoni2022}. While the BH likely spins up through accretion, once the disk becomes magnetically arrested \citep[MAD;][]{Narayan2003,Tchekhovskoy2011}, collapsar BHs will spin down to equilibrium spin of $ a_{\rm eq} \approx 0.1 $ \citep{Jacquemin-Ide2024}. Another pathway for disk formation around slowly spinning BHs is angular momentum loss through magnetic braking of the PNS \citep[e.g.,][]{Duncan1992}. Accretion disks may already form during the PNS phase \citep{Gottlieb2024b}. If the PNS generates outflows that suppress accretion and extend its lifetime beyond the spin-down timescale, the resulting magnetized BH will form with low spin ($a \ll 1$) while still retaining an accretion disk.

As collapsar BHs with $ a \ll 1 $ cannot produce jets via the BZ mechanism, the nature of the transients they might power remains an open question. Recent axisymmetric hydrodynamic simulations \citep{Fujibayashi2023,Fujibayashi2024,Dean2024a,Dean2024b} have explored hydrodynamically-powered outflows from collapsar accretion disks, revealing that they could generate subrelativistic ($ v < 0.1\,c $) outflows with energies up to $\sim 10^{52}\,\erg$. However, collapsar disks may possess significant magnetic flux that could drive faster and more energetic outflows via magnetic-driven flux eruptions \citep[e.g.,][]{Tchekhovskoy2011,Chatterjee2022,Gelles2022,Ripperda2022} and magnetocentrifugally driven jets \citep[BP;][]{Blandford&Payne1982}. This implies that slowly spinning BHs may power unique fast transients, potentially resembling mildly relativistic phenomena such as llGRBs or FBOTs. With the upcoming Rubin Observatory \citep{Ivezic2019}, mapping the full spectrum of collapsar outflows and their optical signatures is becoming increasingly important. These observations could provide crucial insights into collapsar dynamics and shed light on the physics of slowly spinning BHs and collapsar disks.

In this paper, we present the first global 3D general-relativistic magnetohydrodynamic (GRMHD) simulations of a non-spinning BH at the core of a collapsing star. In \S\ref{sec:setup}, we outline the numerical setup, and in \S\ref{sec:diverse} we show how a variety of progenitors give rise to a spectrum of disk-powered outflows. In \S\ref{sec:bp} we analyze the physics of non-spinning BH disk outflows and their observational signatures, focusing on the near-ultraviolet (NUV)/optical signal from cooling emission, which could provide key observational markers for upcoming surveys. In \S\ref{sec:summary}, we summarize our findings and discuss the broader implications for collapsar physics and future observational campaigns.

\section{Numerical setup}\label{sec:setup}

 We perform a series of 3D GRMHD collapsar simulations using the GPU-accelerated code \textsc{h-amr} \citep{Liska2022}, employing an ideal equation of state featuring an adiabatic index of $ \gamma = 4/3 $. The simulations follow the setup detailed in \citet{Gottlieb2022a}, featuring a BH of mass $\mbh = 4\,\msun$ embedded within a Wolf-Rayet star with a stellar radius of $R_* = 4 \times 10^{10}\,\cm $ and mass $M_* \approx 14\,\msun $. We vary the dimensionless spin parameter, $ a $, as outlined in Table~\ref{tab:Simulations}. The only model in which the BH spin might appreciably change during the simulation is $ \alpha0\omega\sigma HR$, which runs for $ 23.3\,\s $ with an initial BH spin of $ a = 0 $. However, we do not evolve the BH spin with accretion, as the outflow physics remains similar between BHs with spins of $ a = 0 $ and $ a_{\rm eq} = 0.1 $, as we will demonstrate. Due to computational time limitations, all other models were run until their outflow type (see below) could be determined. 

The initial mass density profile of the progenitor star is spherically symmetric, defined as
\begin{equation}
     \rho(r) = \rho_0\left(\frac{r}{r_g}\right)^{-1.5}\left(1-\frac{r}{R_*}\right)^3\,,
     \label{eq:Mass Profile}
 \end{equation}
where $\rho_0$ is determined by the condition that $M_* = \int_{0}^{R_*} \rho(r)dV$, and $r_g \equiv G\mbh/c^2 = 6\times 10^5\,\cm $ is the BH gravitational radius. The gas pressure in the star is assumed to be negligible ($ p \ll \rho c^2 $). The specific angular momentum profile of the stellar envelope increases at $r<70\,r_g$, and becomes constant for $ r>70\,r_g $ \citep{Gottlieb2023},
\begin{equation}
     l(r) =
     \begin{cases}
            \omega_0 r^2{\rm sin}^2\theta & r<70r_g \\
            & \\
            \omega_0 \left(70r_g\right)^2{\rm sin}^2\theta & r>70 r_g
     \label{eq:Angular Momentum Profile}
    \end{cases}\,,
\end{equation}
 where $\omega_0 = 5~(50)\,\s^{-1} $ for slowly (rapidly) spinning progenitors that do not form (form) accretion disks.
 
We consider models with and without magnetic fields. For the magnetic vector potential, we assume a uniform magnetic core of radius $ r_c = 10^8\,\cm $ followed by a dipole profile,
 \begin{equation}
     A = A_{\varphi}(r,\theta)\hat{\varphi} = B_0r_c^3\frac{\text{sin}\theta}{r}
      \cdot \text{max}  \left[ \frac{r^2}{r^3+r_c^2}-\left(\frac{R_\star^2}{R_*^3+r_c^3}\right)^3, 0\right] \hat{\varphi}\,.
     \label{eq:Magnetic Vector Potential}
 \end{equation}
We set $B_0$ = $6.4\times 10^{12}\,{\rm G}$ such that the maximum magnetization satisfies: $\sigma_{\rm max} \equiv \text{max}\left(\frac{B^2}{4\pi \rho c^2}\right) \approx 10^{-1.5}$  ; here $ B $ is the comoving magnetic field strength. This choice of magnetic field guarantees that MAD is initiated early on to drive magnetic outflows.
 
\begin{table}
    \setlength{\tabcolsep}{8pt}
	\centering
     \renewcommand{\arraystretch}{1.2}
\label{tab:Simulations}
\begin{tabular}{|c|c c c c c|}
\hline 
    Model name & $a$  & Disk & MAD & Outflow & $t_f\,[\s] $ \\
\hline 
    $a0$ & 0.0  &  No & No & None & 3.5 \\
\hline 
    $a0\omega$ & 0.0 & Yes & No & AS &  1.5 \\
\hline 
    $a0\sigma$ & 0.0 &No & Yes & None &  0.7 \\
\hline 
    $a0\omega\sigma$ & 0.0 & Yes & Yes & APO &  1.5 \\
\hline 
    $a0\omega\sigma HR$ & 0.0 & Yes & Yes & APO &  23.3 \\
\hline 
     $a1\omega\sigma$ & 0.1 & Yes & Yes & APO &  1.4 \\
\hline 
    $a2\omega\sigma$ & 0.2 & Yes & Yes & APO &  1.5 \\
\hline 
    $a3\omega\sigma$ & 0.3 & Yes & Yes & BZ &  1.5 \\
\hline
    $a5\omega\sigma$ & 0.5 & Yes & Yes & BZ &  0.6 \\
\hline
     $a9$ & 0.9 & No & No & None &  2.2 \\
\hline 
     $a9\omega$ & 0.9 & Yes & No & AS &  1.2 \\
\hline 
     $a9\sigma$ & 0.9 & No & Yes & JJ &  0.9 \\
\hline 
     $a9\omega\sigma$ & 0.9 & Yes & Yes & BZ & 1.2 \\
\hline 
\end{tabular}
\caption{Model parameters. The model name structure is as follows: $ a\# $ represents the dimensionless BH spin parameter, where $ \# = 10\times a $; $\omega$ is included (omitted) for models that (do not) form disks with $\omega_0=50\,\s^{-1} $ ($\omega_0=0.5\,\s^{-1} $); $\sigma$ is shown when magnetic fields are included; $HR$ denotes a high-resolution run. The outflow types are accretion shock (AS), accretion-powered outflow (APO), BZ jets (BZ), and jittering jets (JJ). $t_f$ is the final time of the simulation.}
\end{table}

We employ a local adaptive timestep and 2 levels of adaptive mesh refinement (AMR). The grid is spherical, with a logarithmic distribution of cells in the radial direction and uniform distributions in $\hat{\theta}$ and $\hat{\varphi}$  directions. The radial grid extends from just inside the event horizon to $6\times 10^{11}\,\cm $ with numerical resolution at the base AMR level of $N_r\times N_\theta\times N_\varphi$ = $384\times96\times192$ cells, in the r-, $\theta$-, and $\varphi$-directions, respectively. We apply a refinement criterion following \citet{Gottlieb2022c} -- At each radius $r$, the jet and cocoon half-opening angles are measured based on the specific entropy of the fluid. If either half-opening angle contains less than the desired number of cells, $\Delta N_\theta$  = 96 or $\Delta N_\varphi$ = 192, the grid refines to the next AMR level, until it reaches the desired number of cells across each dimension, up to one level of refinement. For the high-resolution model ($a0\omega\sigma HR$), we use up to two levels of AMR.

\section{Diverse outflows}\label{sec:diverse}

We identify several possible outcomes based on the combination of gas angular momentum, which determines whether an accretion disk forms, the magnetic field strength of the disk, which influences the emergence of strongly magnetized outflows, and the BH spin, which governs the power of the BZ jets.

\subsection{Types of outflows}
\subsubsection{No disk: Jittering jets}

\begin{figure}
\includegraphics[scale=0.1]{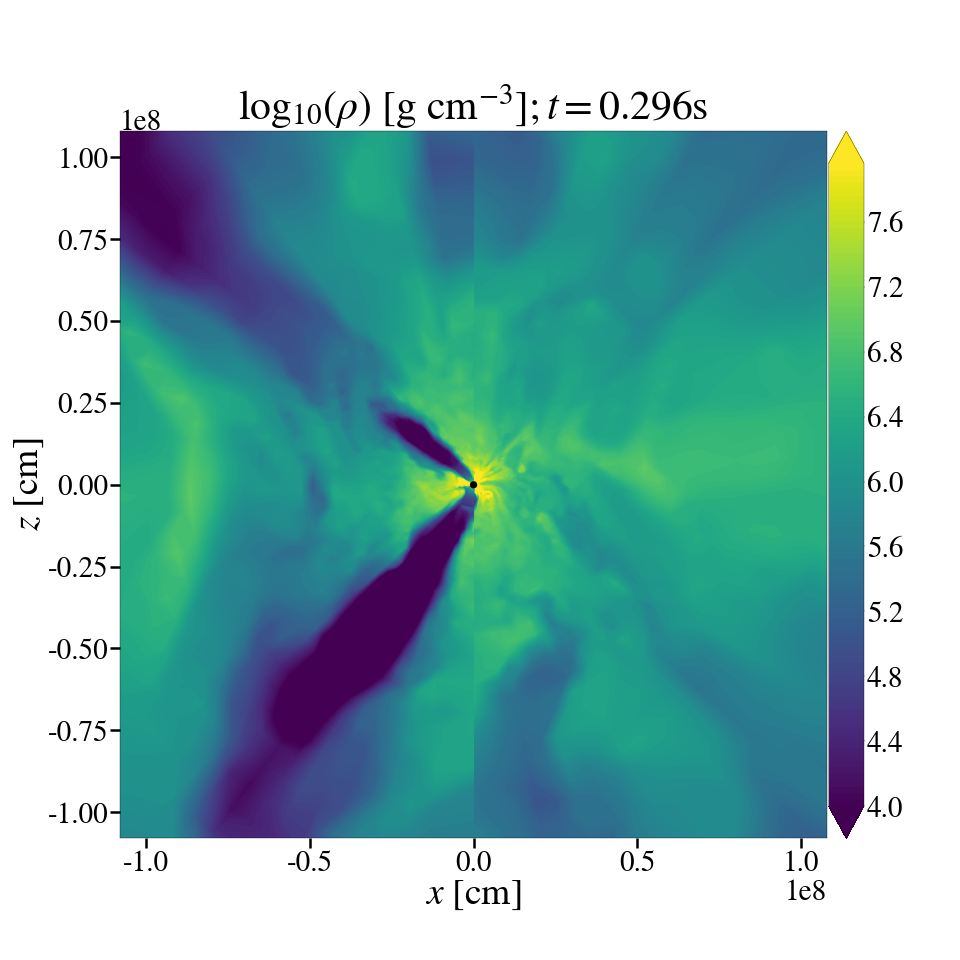}
\caption{Vertical cut of the logarithmic mass density map of model $a9\sigma$ shows that in the absence of an accretion disk, the jets become misaligned, resulting in an intermittent jittering behavior. The lighter blue regions are indicative of previous jittering jet episodes that have mixed with the infalling gas.
\label{fig:Jitteringjet}
}
\end{figure}

Wolf-Rayet progenitor models show increasing angular momentum with radius \citep[e.g.,][]{Woosley&Heger2006}. This suggests that, in some stars, the inner shells may lack sufficient angular momentum to form an accretion disk. However, the BH could still achieve a high spin by accumulating angular momentum from pre-disk accretion. If substantial magnetic flux is present during these early phases, it could, upon reaching the BH, enable the launching of BZ jets (see model $a9\sigma$). However, the absence of an accretion disk has two detrimental effects on the jets: (i) without having the magnetic field lines threading the disk, the field reconnects right away, leading to a prompt jet termination; (ii) Figure~\ref{fig:Jitteringjet} demonstrates that without the angular momentum direction set by the disk, the jets are subject to stochastic angular momentum in the BH vicinity, resulting in a ``jittering jet'' motion that deposits the jets' energy quasi-isotropically in the stellar core, as suggested by \citet{Papish2011,Papish2014,Gilkis2014}.

Over time, the propagation of outflows within a collapsing star can locally enhance angular momentum in certain regions. When these regions reach the BH, they may form an accretion disk. However, this accretion disk will subsequently ingest gas with opposite angular momentum, which destabilizes the disk and the outflows, leading to the formation of a rocking accretion disk \citep[RAD;][]{Lalakos2024}. The RAD formation may gradually align the disk and BH spin, potentially stabilizing the jet and producing more coherent, sustained outflows.

\subsubsection{Magnetized Kerr BHs with disks: BZ-jets (lGRBs)}

In the typical lGRB scenario, a moderately spinning BH, accretion disk, and strong magnetic fields are all required for the BH to launch a BZ-jet, which generates the lGRB (see models $a3\omega\sigma$, $a5\omega\sigma$, $a9\omega\sigma$). In collapsars, BZ-jets are launched nearly maximum (MAD) dimensionless magnetic flux \citep{Gottlieb2023}. Therefore, the jet power is solely governed by the BH spin as
\begin{equation}\label{eq:mad}
    P =\dot{M}\eta(a) c^2\,,
\end{equation}
where $ \eta(a) $ is the jet launching efficiency originating in the BH spin \citep{Lowell2024},
\begin{equation}\label{eq:eta_a}
 \eta(a)=1.063a^4+0.395a^2\,.
 \end{equation}
This highlights the critical role of BH spin in determining the efficiency and power of BZ jet launching in collapsars, and shaping the observable characteristics of lGRBs.

\subsubsection{Disks without a strong magnetic field: Accretion shock}
In scenarios lacking magnetic fields -- likely when the disk is not initially present to preserve the PNS's magnetic field \citep{Gottlieb2024b} -- the disk transports angular momentum outward through hydrodynamic viscous forces. This process energizes an expanding accretion shock (see models $a0\omega$, $a9\omega$). While such shocks are also present in magnetically driven outflows, their contribution is more prominent when magnetic forces are absent. The shock exhibits an $m = 1$ mode, similar to accretion shocks observed in core-collapse SN (CCSNe) simulations \citep{Blondin2003,Blondin2006}.

\subsubsection{Slowly spinning BHs with magnetic disks: Accretion-powered outflows}
Slowly spinning BHs lack the rotational energy needed to efficiently launch BZ jets, allowing other types of outflows to dominate. In the presence of a magnetic accretion disk (see models $a0\omega\sigma$, $a0\omega\sigma HR$, $a1\omega\sigma$, $a2\omega\sigma$), the disks drive magnetically-dominated collimated outflows that ultimately lead to the star's explosion. We investigate the properties of these outflows in \S\ref{sec:bp}.

\subsection{Outflow comparison}

\begin{figure}
\includegraphics[scale=0.1]{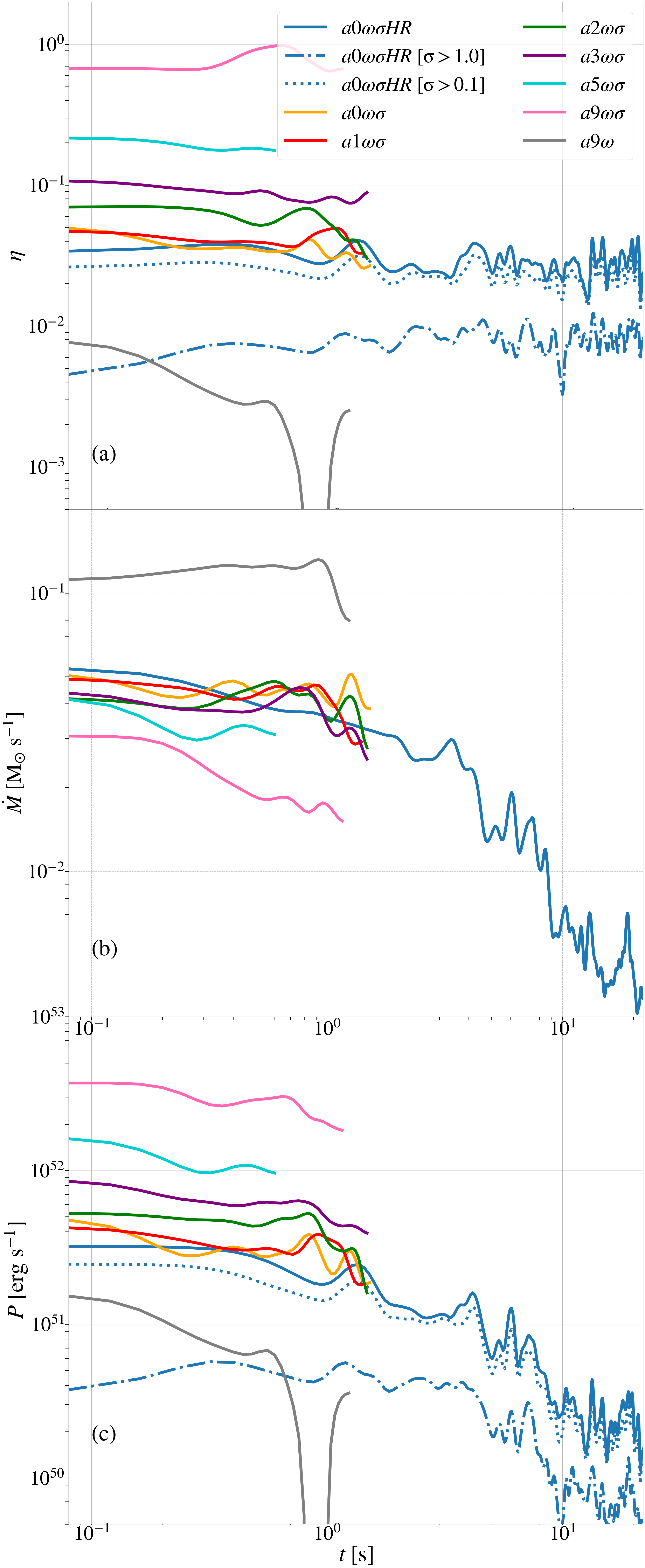}
\caption{Smoothed time evolution at the horizon for models with fast rotation that form accretion disks. Magnetic disks eject outflows more efficiently than hydrodynamic disks (model $ \alpha9\omega $). Higher BH spins increase the launching efficiency (panel a), resulting in more powerful outflows (panel c). Consequently, higher spins exert stronger feedback on accretion, moderately reducing the mass accretion onto the BH (panel b).
\label{fig:L_ETA_Mdot}
}
\end{figure}

\begin{figure}
\includegraphics[scale=0.1]{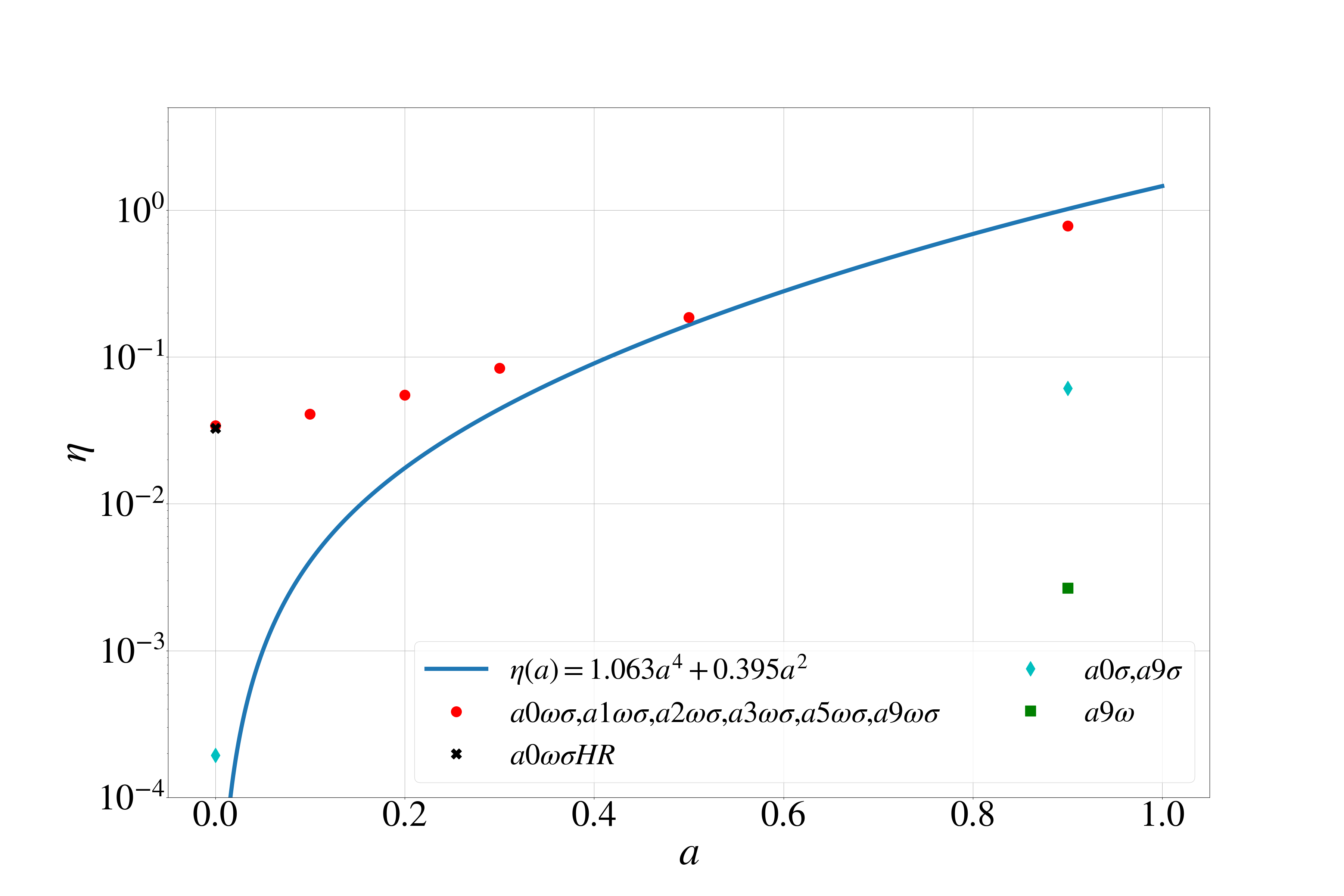}
\caption{The time-averaged outflow efficiency in simulations with strong magnetic fields and rapid rotation (red dots) follows Eq.~\eqref{eq:eta_a} (blue curve) for $ a \gtrsim 0.3 $. For $ a \lesssim 0.3 $, the outflow efficiency is dominated by disk outflows rather than the BH. BHs with weak magnetic fields (green square) or without accretion disks (turquoise diamonds) exhibit a low launching efficiency.
\label{fig:EfficiencyvSpin}
}
\end{figure}

Figure~\ref{fig:L_ETA_Mdot} shows the time evolution of various quantities at the BH horizon in simulations that form accretion disks. Panel (a) of Figure 2 compares the outflow launching efficiency, defined as
\begin{equation}
    \eta = \frac{P}{\dot{M}c^2}\,,
     \label{eq:Jet Efficiency}
\end{equation}
where the power of the outflows is defined as
\begin{equation}\label{eq:power}
    P = \int_{r_g} \sqrt{-g} (-T^{r}_{t} - \rho u^r) d\theta d\varphi\,,
\end{equation}
with $ g $ being the metric determinant, $ T^{r}_{t} $ denotes the radial energy flux density component of the mixed stress-energy tensor, $ u^\mu $ is the four-velocity such that $ \rho u^r $ represents the radial mass flux density, and the integration is done over the BH horizon. As discussed in \S\ref{sec:setup}, the magnetic disk remains in a MAD state throughout. Thus, for a rapidly spinning BH which drives BZ jets, the launching efficiency is governed by Eq.~\eqref{eq:mad}, while for a low spin, the efficiency is dominated by disk outflows, as we show in Figure~\ref{fig:EfficiencyvSpin}. The decomposition of the disk outflow’s launching efficiency in model $ \alpha0\omega\sigma HR $, based on plasma magnetization (blue lines), indicates that the emission is magnetically driven, governed by a dynamically important magnetic field with $ \sigma \gtrsim 0.1 $ (dotted lines). The mechanisms driving these outflows are discussed in \S\ref{sec:bp}.

Panel (b) of Figure 2  shows the mass accretion rate, $ \dot{M} $, onto the BH across different models, demonstrating that the strong outflows, produced by higher spin BHs, slightly suppress accretion. $a0\omega\sigma HR$ exhibits a sharper decrease in accretion at $ t \gtrsim 0.5\,\s $ due to this feedback mechanism. Panel (c) of Figure 2 depicts the radial power at the horizon [Eq.~\eqref{eq:power}], indicating total outflow energies of $ E \sim 10^{51}-10^{52}\,\erg $.

Figure~\ref{fig:EfficiencyvSpin} compares the empirical expression for the spin efficiency of BZ jets in Eq.~\eqref{eq:eta_a} (blue curve) with the time-averaged launching efficiency of various outflows in our simulations. For each spin, the maximum power is achieved in simulations with both strong magnetic fields and fast rotation (red dots and black X). The empirical equation is consistent with the numerical values of simulations with $ a \gtrsim 0.3 $, implying that the BZ mechanism dominates the outflows. Simulations of magnetic disks with $a\lesssim 0.3$ exhibit outflows with considerably higher efficiency than the prediction by the blue curve. This indicates that an additional ejection mechanism is at play, driven by the accretion disk. For example, when $ a = 0 $, the efficiency $ \eta \approx 3\% $ is equivalent to BZ outflows from a BH with spin $ a_{\rm crit} \approx 0.25 $ (corresponds to $ \eta(a) \approx 3\% $). This suggests that outflows from BHs with $ a \lesssim 0.25 $ are dominated by non-BZ outflows, consistent with \citet{McKinney2012,Narayan2012,Tchekhovskoy2012}. Simulations without rotation (turquoise diamonds) prevent disk formation, leading to brief, jittering jet launches due to the accumulation of flux on the BH. When magnetic flux is absent (green square), the resulting hydrodynamic outflows are weaker by orders of magnitude compared to the magnetically driven outflows.

\begin{figure*}
\centering
{\includegraphics[scale=0.073]{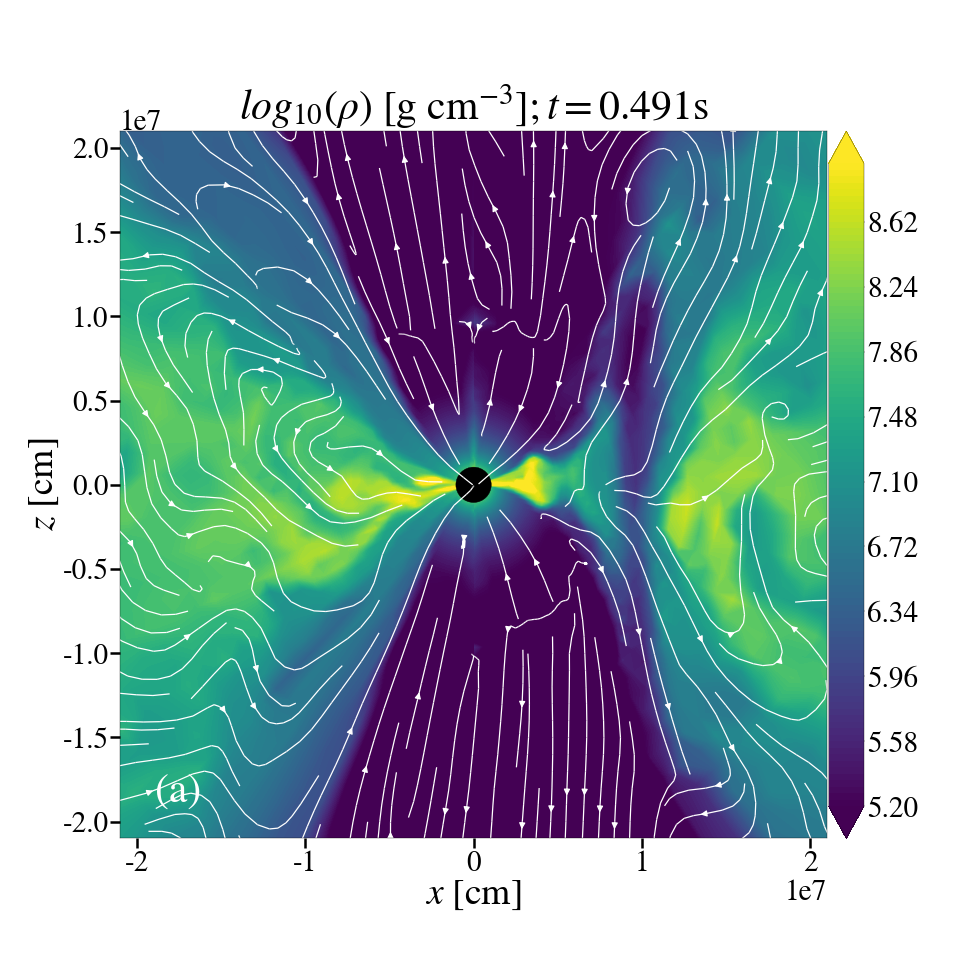}}
{\includegraphics[scale=0.073]{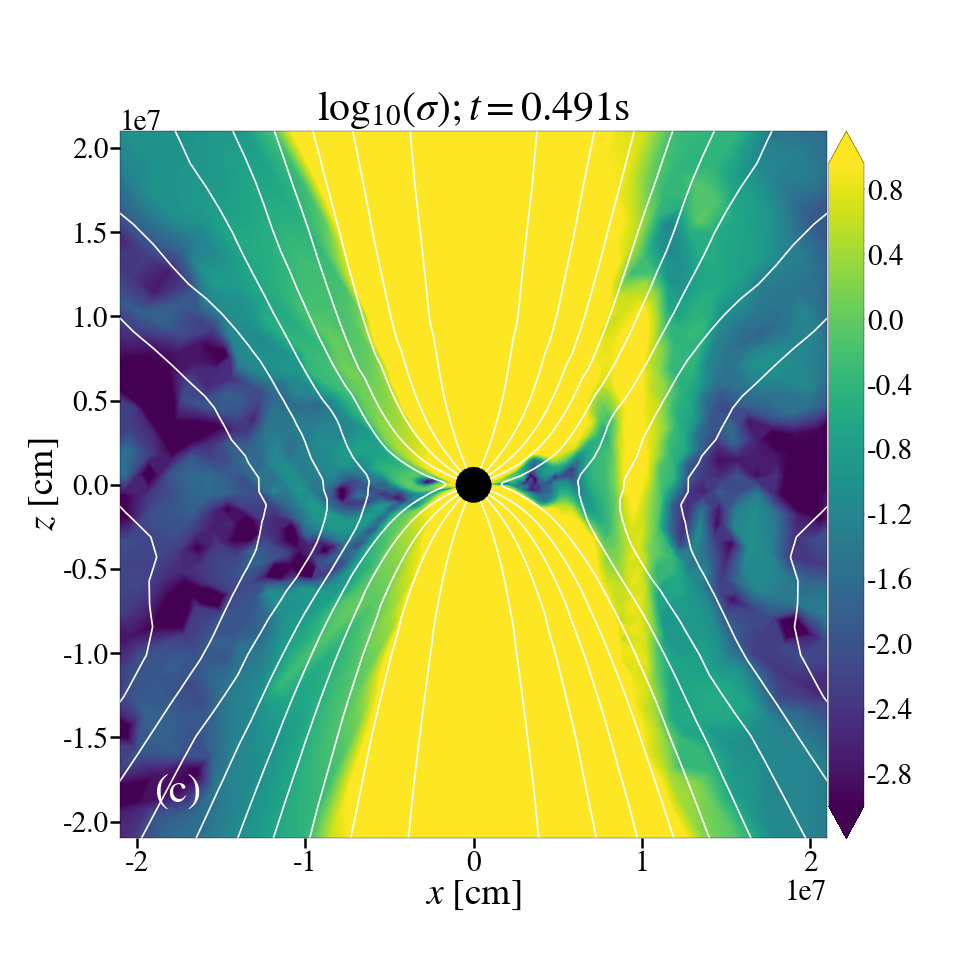}}
{\includegraphics[scale=0.073]{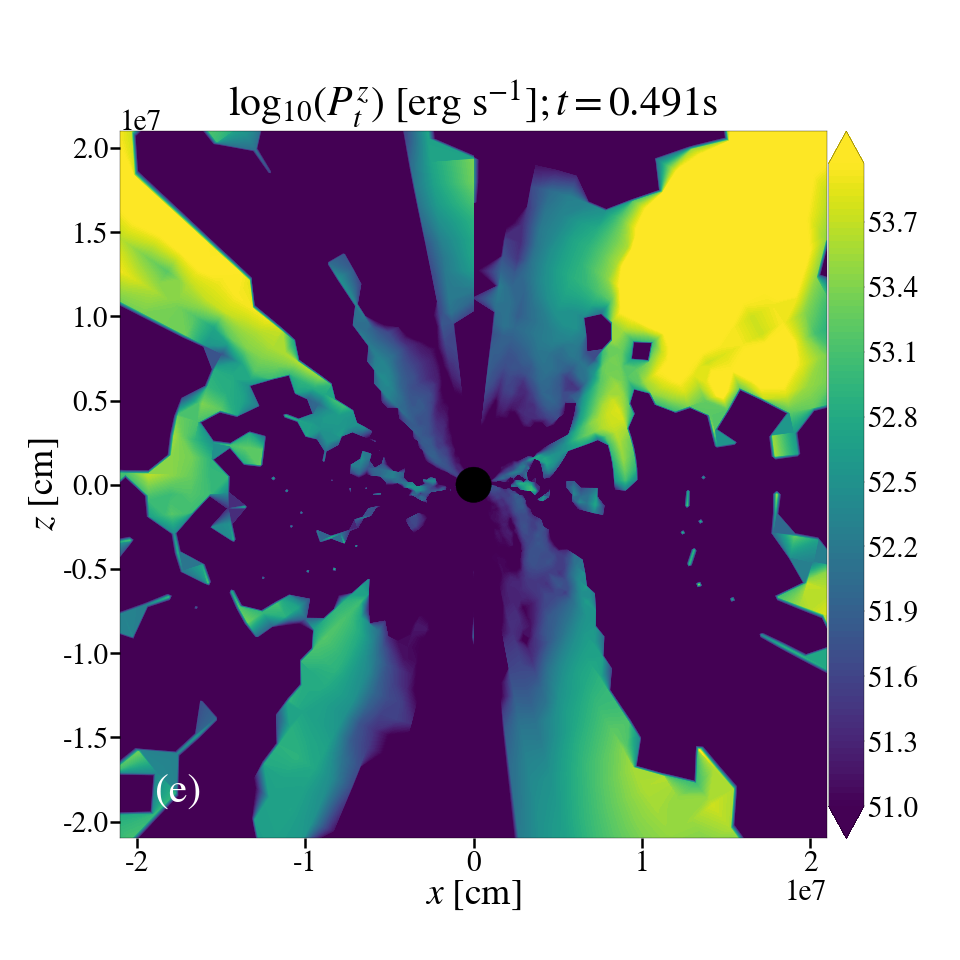}}
{\includegraphics[scale=0.073]{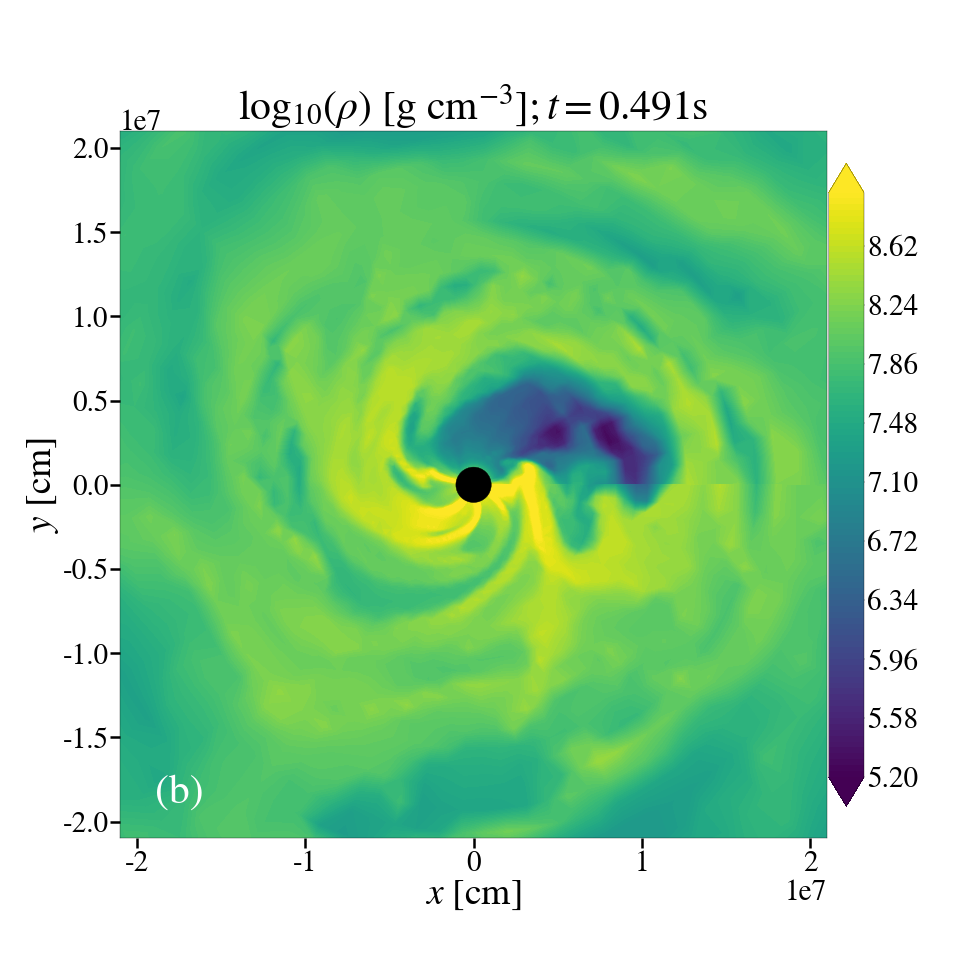}}
{\includegraphics[scale=0.073]{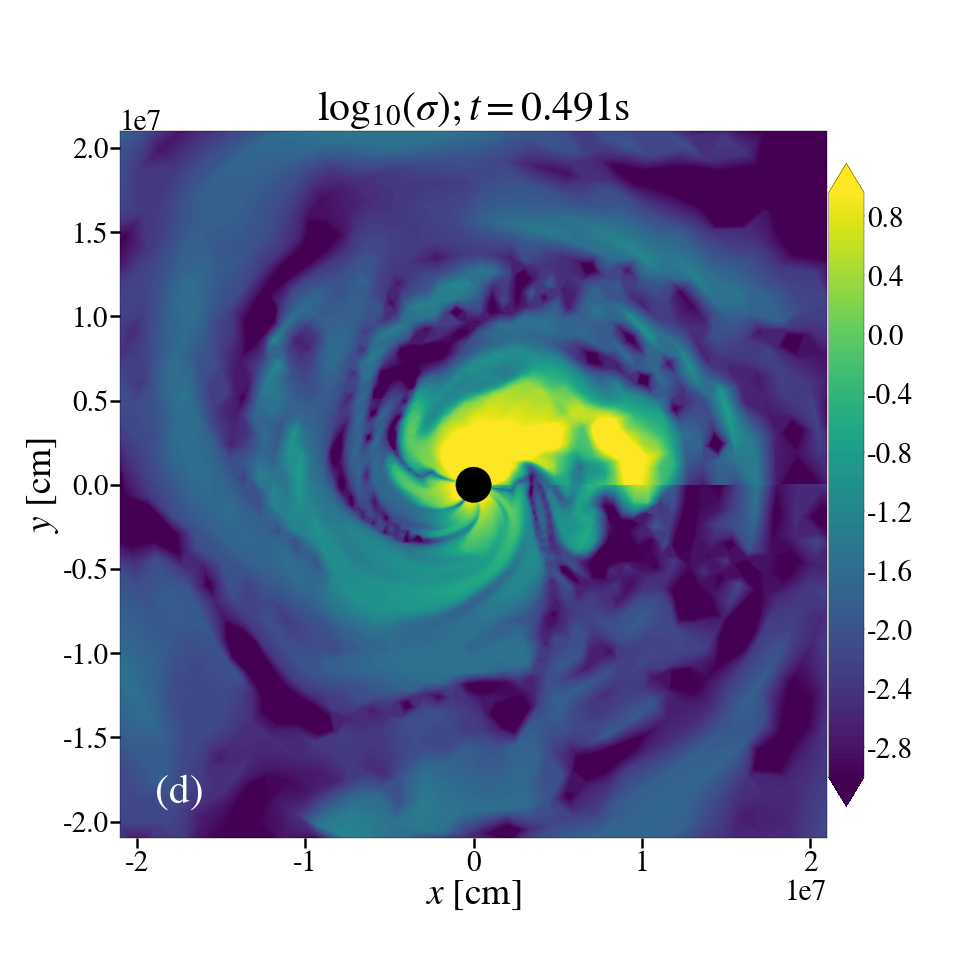}}
{\includegraphics[scale=0.073]{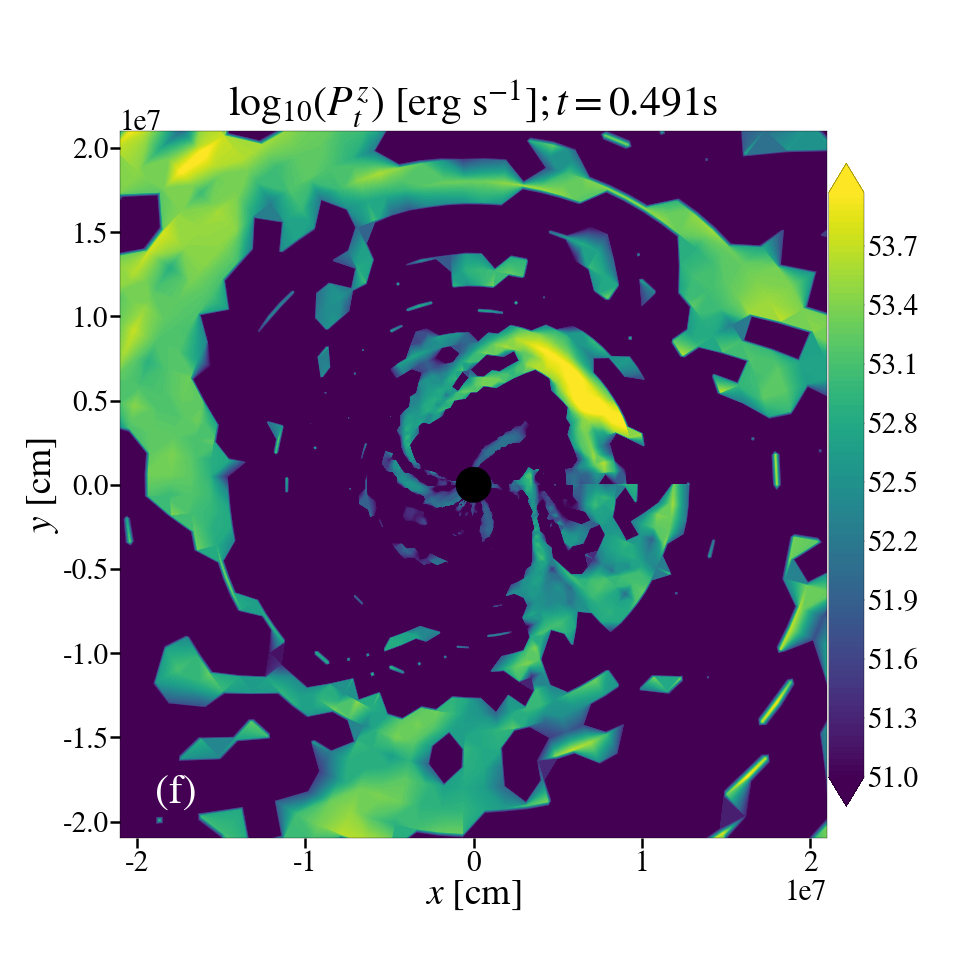}}
\caption{Meridional (top) and equatorial (bottom) maps at the time of a flux eruption ($ \hat{y}>0 $) in the innermost $ \sim 2\times10^7\,\cm $. The logarithm of the mass density [panels (a) and (b)] and magnetization [panels (c) and (d)] maps show the buildup of magnetic flux and the low-density region associated with it in the disk (bottom panels), leading to vertical outflows (top). The white arrows in panel (a) represent velocity streamlines, while the white lines in panel (c) represent magnetic field lines. Panels (e),(f) display the power of the outflows in the $ \hat{z} $ direction, demonstrating that the bulk of the flux originates from the eruption. We note that the high magnetization connected to the non-spinning BH in panel (c) contains no energy, and originates from the floor values in the code.
\label{fig:MAPS}
}
\end{figure*}

\section{Disk-Powered Outflows}\label{sec:bp}

We analyze the magnetized non-spinning BH explosion model with a higher resolution ($ a0\omega\sigma HR $) to investigate the driving mechanism of the outflows. To examine the observational signatures of the resulting ejecta, we continue the simulation well beyond breakout at $ t_b \approx 11\,\s $ until $ t = 1.16 \times 10^6\,r_g/c \approx 23.3\,\s $.

\subsection{Explosion physics}

Magnetic disks can exhibit at least two types of outflows:

(i) Steady collimated outflows (BP jets): when poloidal magnetic field lines thread a rotating disk, magnetocentrifugal launching accelerates the plasma along the twisted stretched field lines, driving material to be expelled along the field lines. This process launches collimated magnetocentrifugal winds.

(ii) Magnetic flux eruptions: when the magnetic field saturates on the BH, the high magnetic pressure gradient near the BH winds the magnetic field lines via differential rotation in the disk. As the plasma is transported radially outward from the BH’s magnetosphere, it relaxes the field by magnetic reconnection and plasma eruption from the disk.

\begin{figure*}[]
\centering
\includegraphics[width=\textwidth,height=0.62\textheight,keepaspectratio]{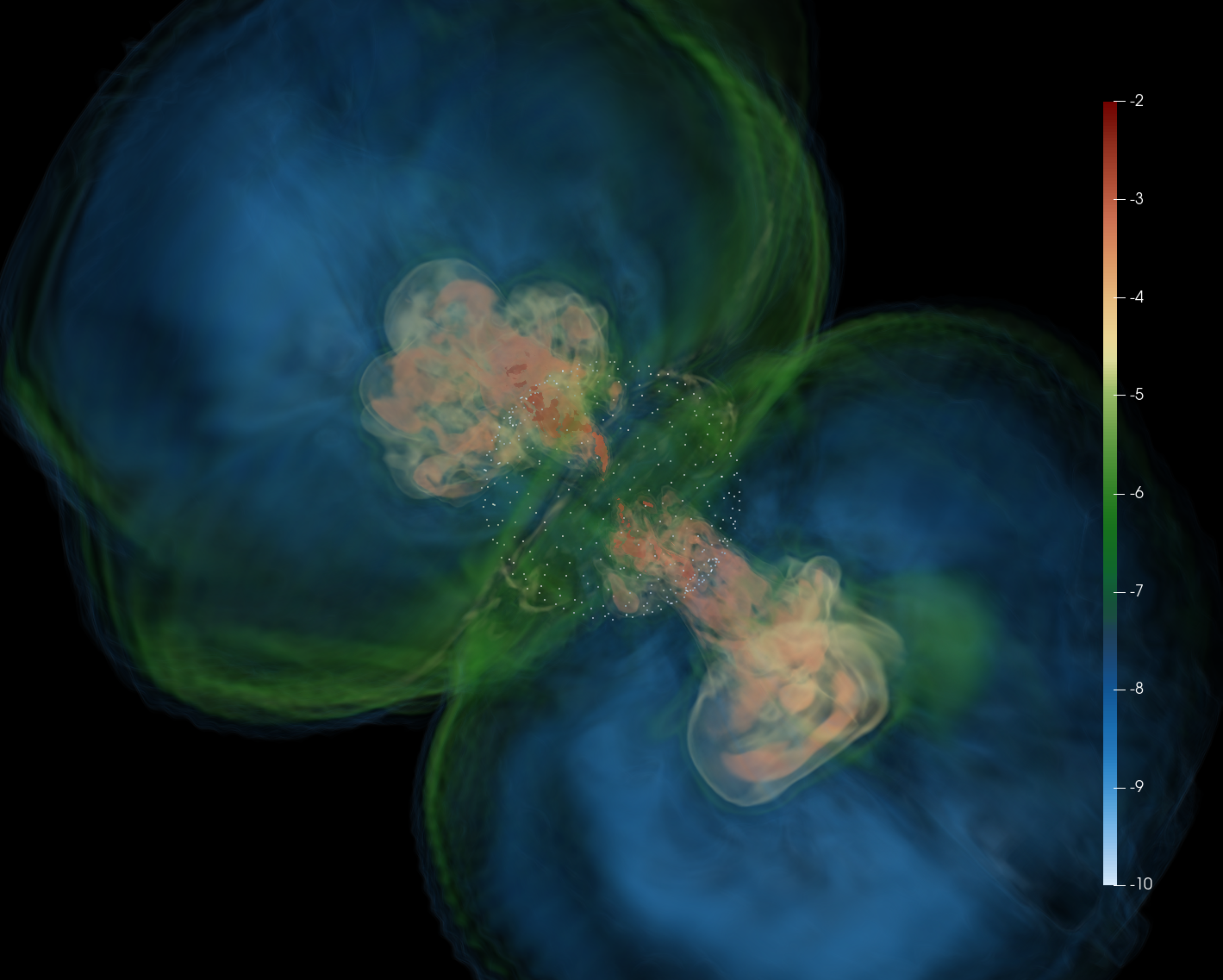}
\caption{3D rendering of the logarithm of the magnetization in model $a0\omega\sigma HR$ at $t = 23.3 \,\s $. Massive outflows moving at $ v \approx 0.15c $ (red) unbind faster, lighter ejecta traveling at $ v \lesssim 0.4c $, forming a double-lobed structure (blue). The star's original radius is outlined by the white dotted sphere.
\label{fig:0spin_end}
}
\end{figure*}

Fig.~\ref{fig:MAPS} shows meridional (top row) and equatorial (bottom row) maps of the region near the BH during a magnetic flux eruption in the non-spinning BH model $a0\omega\sigma HR$. Magnetic pressure accumulates in low-density regions near the BH [panels (b) and (d)] and is transported radially outward. Ultimately, the plasma is expelled from the disk [see velocity streamlines in panel (a)] along the vertical field lines [white lines in panel (c)]. In addition to the collimated flux eruptions, magnetically driven winds are constantly launched along thepoloidal field lines as BP outflows \citep[see e.g.,][]{Musolino2024}. However, during the flux eruptions, most of the vertical power of the outflow,
\begin{equation}
    P^z_t =  r^2\sqrt{g_{rr}}T^{r}_{t}\text{cos}\theta +  r^2\sqrt{g_{\theta \theta}}T^{\theta}_{t}\text{sin}\theta \,,
\end{equation}
is concentrated in the eruptions, as shown by the vertical flux maps in panels (e) and (f).
 
Fig.~\ref{fig:MAPS}(e),(f) demonstrates that during an eruption, the vertical power from the flux eruption constitutes almost all the disk power. This indicates that the integrated energy from these flux eruptions may dominate over the BP mechanism, making flux eruptions the primary driver of the disk outflows.

\begin{figure}
\includegraphics[scale=0.1]{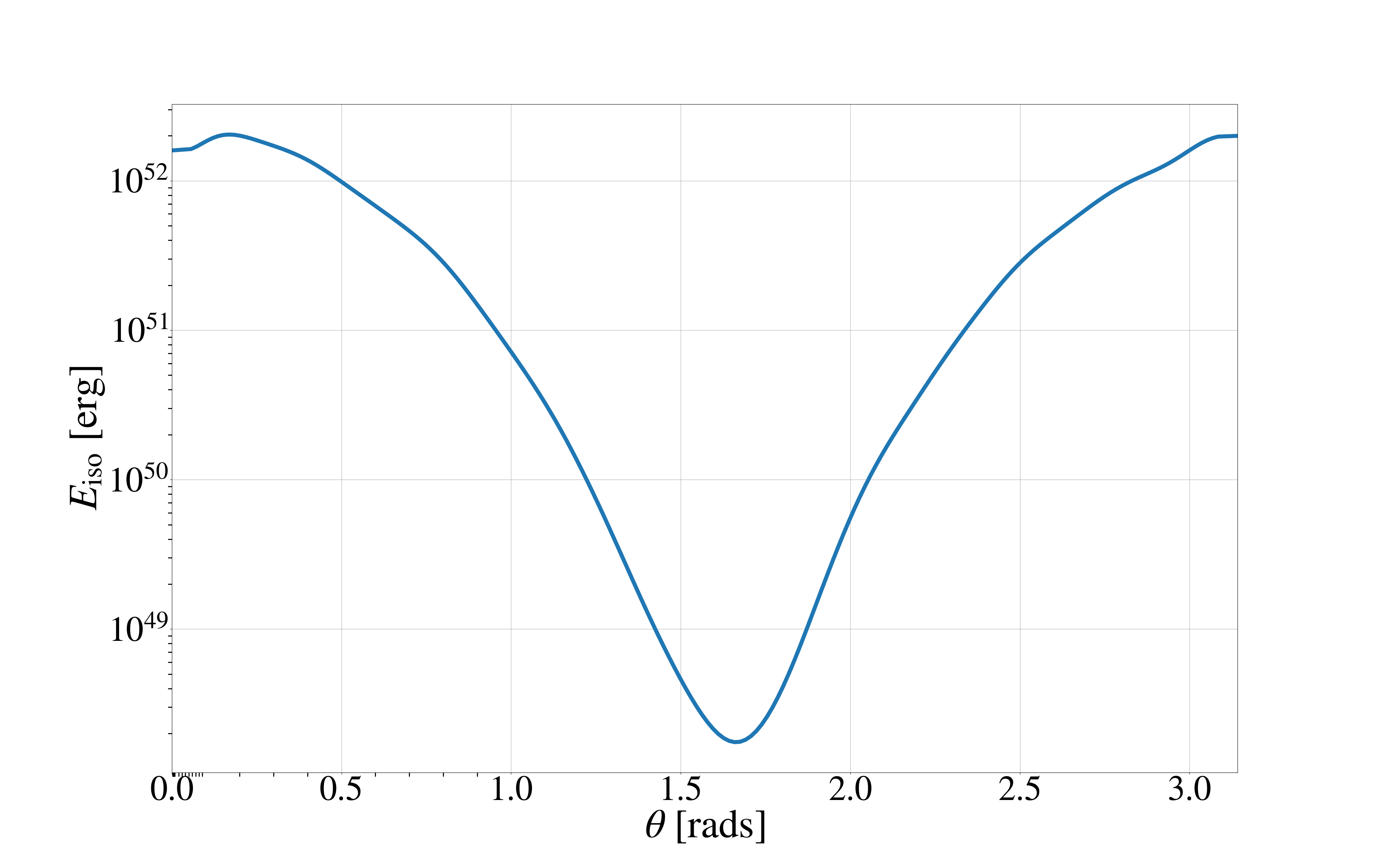}
\includegraphics[scale=0.1]{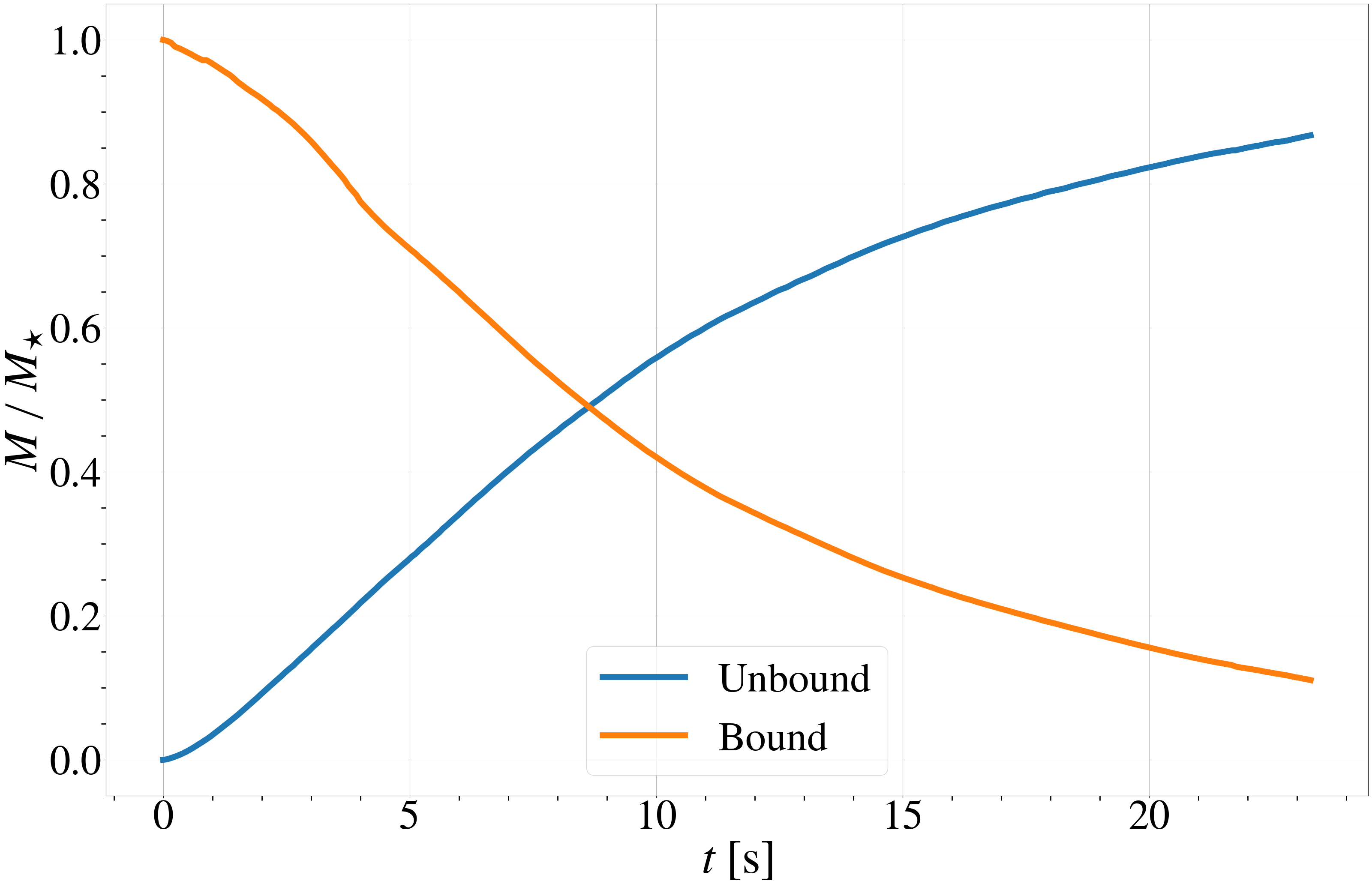}
\includegraphics[scale=0.1]{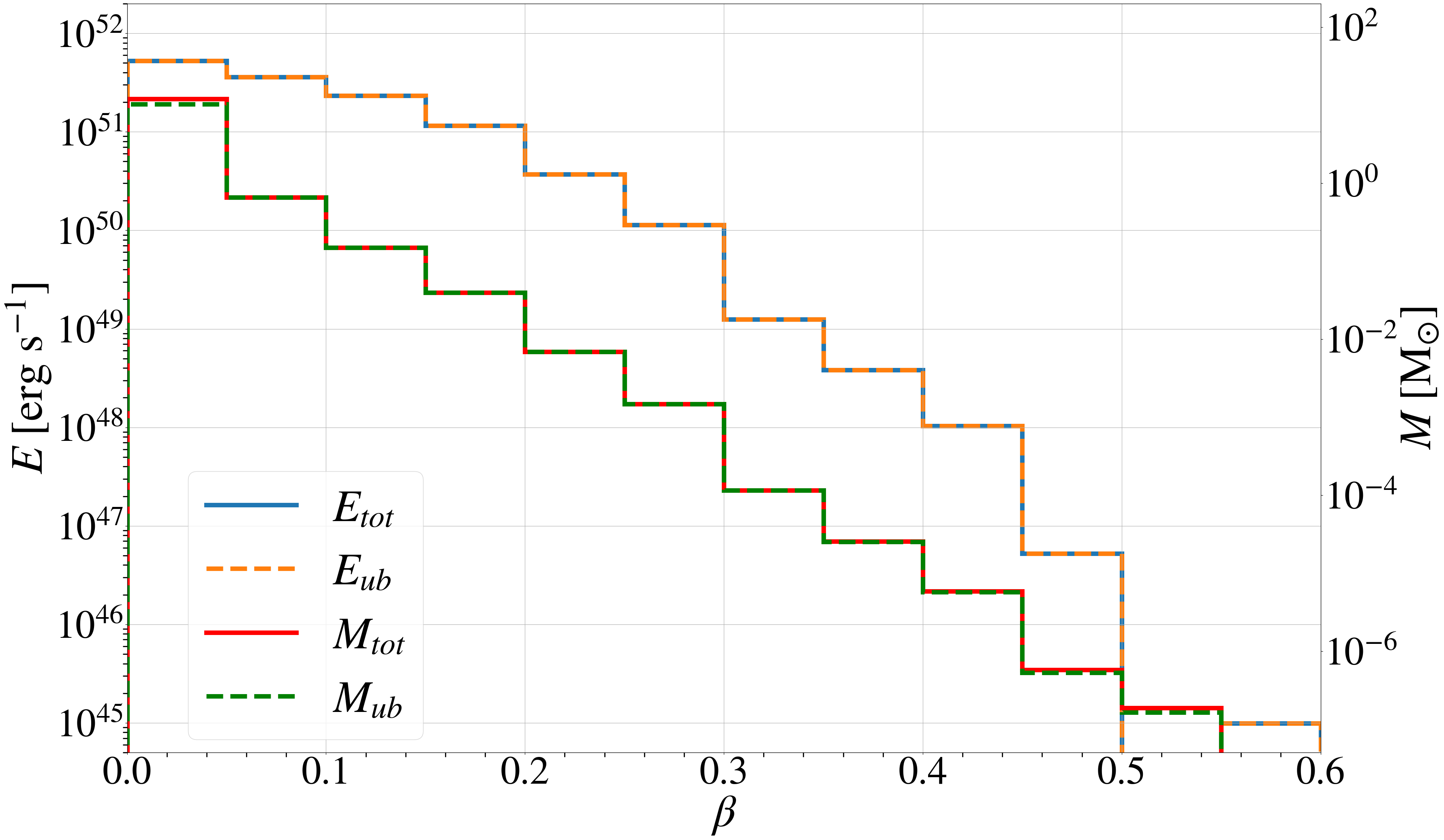}
\caption{Distributions for model $ \alpha 0\omega\sigma HR $. Top panel: Angular distribution of the isotropic equivalent energy of the gas at $ r > R_\star $ shows that the polar outflows possess $ E_{\rm iso} \approx 10^{52}\,\erg $ at $ t = 23.3\,\s $. Middle panel: Bound (orange) and unbound (blue) mass distributions show that the disk outflows unbind most of the star, leaving only a small fraction to supply for the BH mass growth. Bottom panel: Histograms of the (unbound) mass (red) and energy (blue) as a function of $\frac{v}{c}$ at $t=23.3\,\s$ reveal a large amount of energy that is uniformly distributed at $ \beta < 0.3 $.
}
\label{fig:hist}
\end{figure}

Figure~\ref{fig:0spin_end} presents a 3D rendering of the magnetization of the ejecta after breakout from the star (white dotted sphere). The mildly collimated, massive disk outflows (red) at $ v \approx 0.15c $ are embedded within a quasi-spherical, shocked, double-lobed structure (blue) moving at $ v \lesssim 0.4c $. As the disk-driven outflows interact with the infalling stellar material, they form a double-shocked cocoon layer. The shocked cocoon material falls back onto the disk, tilting it away from the axis of rotation and driving the wobbling motion. Our temporal resolution has prevented us from conducting a detailed analysis of the wobbling origin \citep[see][for analysis and discussion]{Gottlieb2022c}. The top panel of Figure~\ref{fig:hist} depicts the angular distribution of the isotropic equivalent energy of the outflows outside the star. It illustrates that the $ E \approx 10^{51.5}\,\erg $ ejecta is concentrated within a characteristic opening angle of $ \theta_{e} \approx \pi/4 $, resulting in isotropic equivalent energy of $ E_{\rm iso} \approx 10^{52}\,\erg $.

The middle panel of Fig.~\ref{fig:hist} shows the time evolution of the bound (orange) and unbound (blue) mass in model $a0\omega\sigma HR$, determined by the Bernoulli parameter criterion $-(h + \sigma)u_{t}>1$, where $h = 1 + 4p/ \rho c^2$ is the specific enthalpy, and $u_t$ is the covariant time component of the four-velocity vector. The wobbling of the system causes the outflows to expand in multiple directions, misaligning with the spin axis and increasing the effective interaction area with the collapsing star. As a result, the disk outflows unbind $ \gtrsim 90\% $ of the star, leading to a substantial SN ejecta. The bound mass, comprising $ \lesssim 10\% $ of the star, indicates that the BH mass will remain relatively close to its birth mass, likely falling within the ``mass gap''.

The bottom panel of Fig.~\ref{fig:hist} displays the binned mass (red) and total energy (blue) distributions of the gas as a function of the dimensionless velocity $\beta = v/c $. The unbound mass satisfies $ M_{\rm ub} \sim 10^{-10\beta} $, with energy displaying a similar monotonic decline with $ \beta $. This decaying energy distribution contrasts with those associated with the non-relativistic portion of the cocoon formed by the interaction between relativistic jets and the star \citep{Gottlieb2021,Eisenberg2022}. This difference suggests that the observational signatures, discussed in the following section, likely vary between disk-powered and jet-cocoon-powered sub-relativistic outflows.

\subsection{Electromagnetic signatures}\label{sec:em}

The large amount of energy in the disk outflows indicates that they may generate various detectable electromagnetic signals. The first light will emerge as the mildly relativistic shock breaks through the star, producing an X-ray/UV shock breakout. As the gas expands adiabatically, the emission will shift toward softer wavelengths, powering a cooling NUV/optical signal. Once the bulk of the ejecta reaches the photosphere after a few weeks, it may power $^{56}$Ni decay-powered emission. Finally, the interaction between the disk outflows and the circumstellar medium (CSM) will accelerate CSM electrons, generating a synchrotron-powered afterglow. In this section, we provide a rough estimate of the cooling emission, neglecting potential effects from ejecta--CSM interaction \citep[see e.g.,][]{Matsumoto2022}. Detailed calculations of this interaction and other expected signals are left for future work.

The slowest gas reaching homologous expansion at the end of our simulation is located at $ r \approx R_\star $ with $ \beta_{\rm min} \approx 0.05 $. We calculate the emission for shells moving at $ \beta > \beta_{\rm min} $ using the gas properties at $ t = t_f $. For each bin with mass $ m $ and velocity $ \beta $, the \emph{observed} emission time of the luminosity shell is set by the time it takes the shell to reach $ \tau = 1/\beta $ \citep[see e.g.][]{Gottlieb2023d},
\begin{equation}
    t_{\rm obs} \approx \sqrt{\frac{m\kappa\Gamma(1-\beta)^3}{4\pi \beta c^2}}\,,
     \label{eq:Time cool}
\end{equation}
where $ \Gamma $ is the Lorentz factor of the shell, and we use electron scattering opacity $\kappa = 0.2\,\cm^{2}\,{\rm g}^{-1}$ for the plasma.

The observed emission at time $ t_{\rm obs} $ from the luminosity shell moving at $ \beta $ is estimated by the radial thermal energy flux of the shell,
\begin{equation}
    L \approx -\int_{r=\beta t_f} \sqrt{-g} \gamma \epsilon u^r u_t(1-\beta)^{-4}\left(\frac{t_f}{t_{\rm obs}}\right)^2 c^3 d\theta d\varphi\,,
     \label{eq:E th}
\end{equation}
where $ \epsilon $ is the thermal energy density, $ (1-\beta)^{-4} $ is the boost from the lab frame to the observer frame, and $ \left(t_f/t_{\rm obs}\right)^2 $ accounts for adiabatic losses of the flux carried by the shell.

At the relevant times of emission, the emitting shells are subrelativistic, allowing enough time for the gas to cool down and thermalize so that the temperature can be approximated by a blackbody,
\begin{equation}
    T \approx \left[\frac{L}{4\pi\sigma_{\rm SB} (\beta c t_{\rm obs})^2}\right]^{1/4}\,,
     \label{eq:Temp}
\end{equation}
where $ \sigma_{\rm SB} $ is the Stefan-Boltzmann constant.

Using the luminosity and temperature, we determine the spectral luminosity,
\begin{equation}
    L_{\nu}(\nu) = \frac{(1-\beta)L}{1.53\nu_{max}}\frac{e^{\frac{h\nu_{max}}{kT}}-1}{e^{\frac{h\nu }{kT}}-1}\left(\frac{\nu}{\nu_{max}}\right)^3\,,
     \label{eq:L nu}
\end{equation}
where $h$ is the Planck constant, $k$ is Boltzmann constant, and $h\nu_{max}= 2.821\,kT$.

\begin{figure}
\includegraphics[scale=0.1]{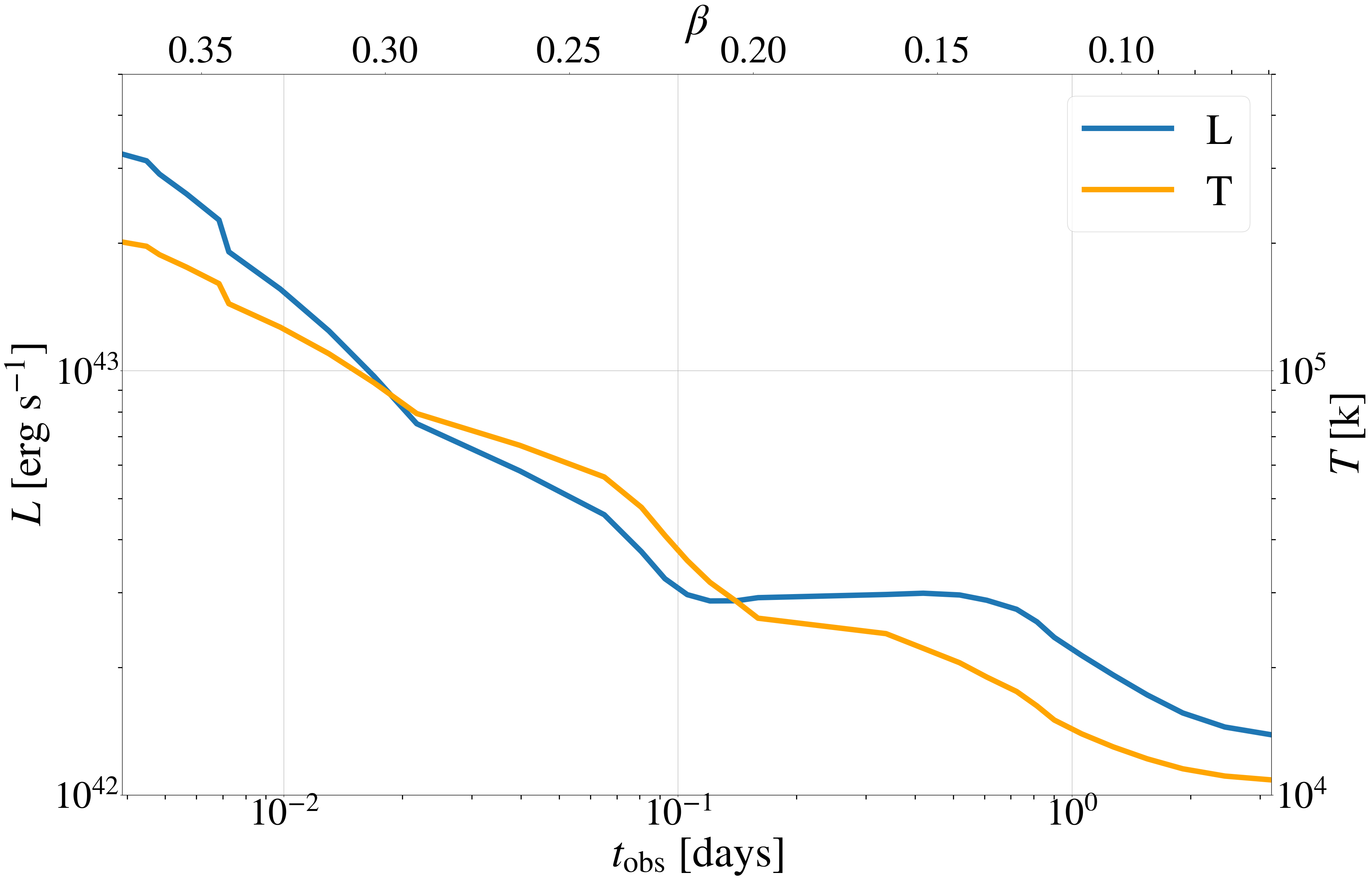}
\includegraphics[scale=0.1]{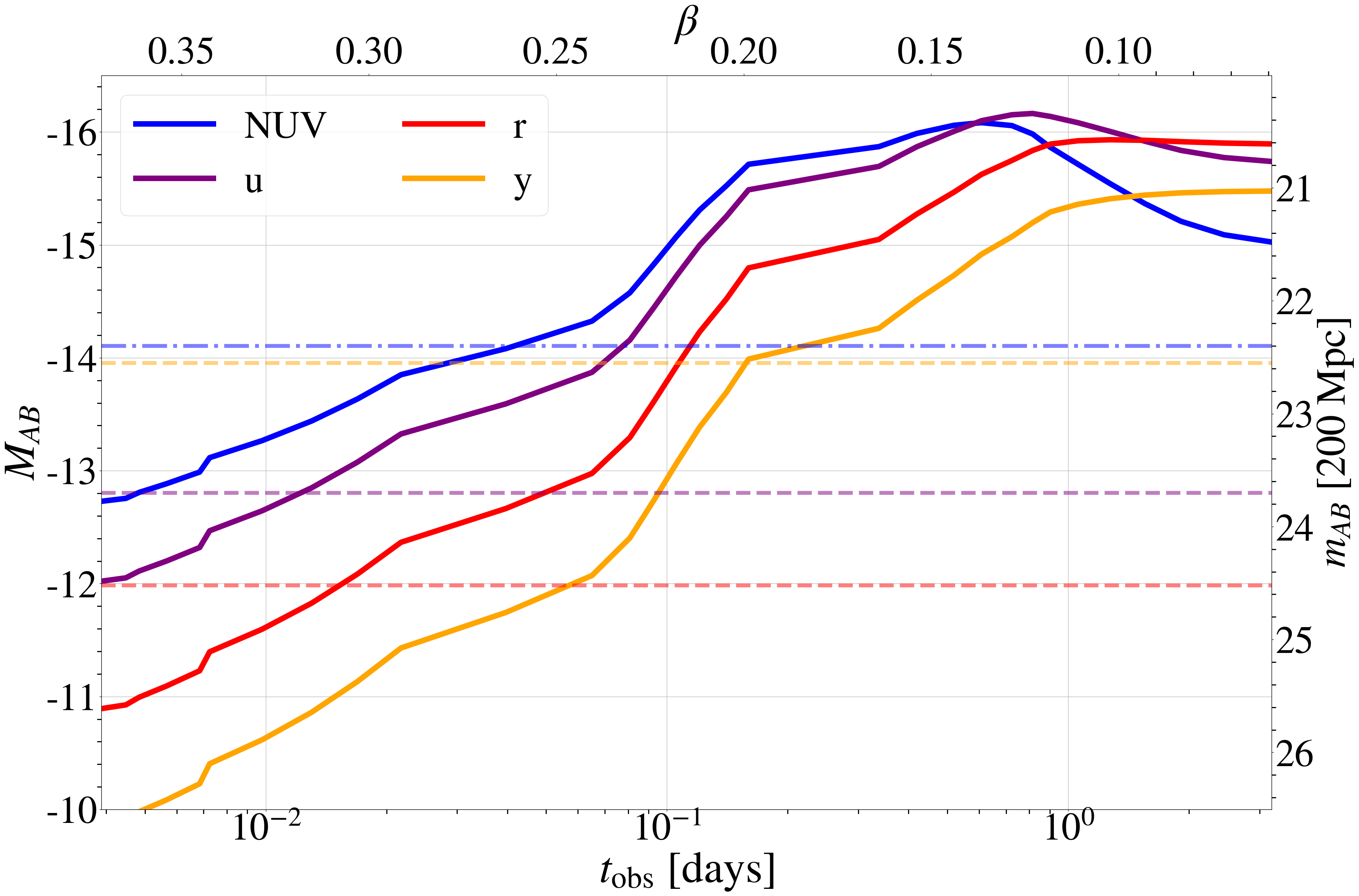}
\caption{Top: The bolometric luminosity (blue) and temperature (orange) as functions of observed time (bottom horizontal axis) and emitting shell velocity (top horizontal axis). Bottom: Absolute (left vertical axis) and apparent (right vertical axis) AB magnitudes at $ 200\,{\rm Mpc} $ in various UV/optical bands. The dashed lines indicate the approximate $ 5\sigma $ point source limiting magnitudes for the Rubin Observatory bands: u ($8.4 \times  10^{14}\, {\rm Hz}$), r ($4.9 \times  10^{14}\, {\rm Hz}$), y ($3.0 \times  10^{14}\, {\rm Hz}$), and ULTRASAT/UVEX NUV ($1.2 \times  10^{15}\, {\rm Hz}$) band.
\label{fig:light_curves}
}
\end{figure}

The top panel of Figure~\ref{fig:light_curves} shows the bolometric luminosity (blue) and temperature (orange). The similar decline in luminosity and temperature leads to a fast rise in the NUV/optical bands, which fall in the Rayleigh-Jeans tail of the spectrum. The bottom panel of Figure~\ref{fig:light_curves} illustrates the time evolution of the NUV/optical bands during the first few days. The NUV peaks at $ M_{\rm AB} \sim -16 $ after $ \sim 1\,{\rm day} $, while the optical emission likely persists for $ \sim 1\,{\rm week} $\footnote{Our estimates extrapolate the homologous gas at $ \beta > \beta_{\rm min} $, corresponding to $ t \lesssim 3\,{\rm days} $. Capturing the later evolution of the gas will require a longer simulation.}.

Upcoming optical survey Vera C. Rubin Observatory \citep{LSST2009,Ivezic2019} and UV satellites such as the Ultraviolet Transient Astronomy Satellite \citep[ULTRASAT;][]{Sagiv2014,Ben-Ami2022,Shvartzvald2024} and Ultraviolet Explorer (UVEX; \citealt{Kulkarni2021}) will survey the sky with a cadence of a few days, comparable to the duration of these signals. The dashed lines in the bottom panel of Figure~\ref{fig:light_curves} represent the approximate $5\sigma$ point source limiting magnitude at $ 200\,{\rm Mpc} $ that will be detectable in various facilities. The sensitivity lines indicate that these sources can be detected out to a distance of hundreds of Mpc by ULTRASAT, and to $ D > {\rm Gpc} $ by UVEX and Rubin Observatory.

ULTRASAT will observe $ \sim 6800$ deg$^{2}$ of the sky during its low-cadence survey mode, corresponding to $f_{\Omega} \approx 1/6 $ of the sky. If all Type Ic-BL SNe harbor such accretion disks, and assuming emission only within $ \theta_e $, then the ULTRASAT detection rate is $ \sim (2\pi/3)\theta_e^2 D^{3}\mathcal{R}_{\rm Ic-BL}f_{\Omega} \sim 30\,{\rm yr}^{-1} $. A similar calculation for Rubin Observatory yields a detection rate of $ \gtrsim 10^4\,{\rm yr}^{-1} $. However, these rates strongly depend on the disk properties such as magnetization, and the uncertain rate of disk formation in Type Ic-BL SNe and other CCSNe classes.

\section{Summary and Discussion}\label{sec:summary}

Most BHs formed from massive star collapse are slow-spinning. Accretion disks around these BHs are likely to form through at least two pathways. First, the large angular momentum reservoir at the outer radii of collapsing stars suggests late disk formation around a slowly spinning BH. Alternatively, if the spin-down time of an accreting PNS exceeds the PNS's collapse time into a BH, the PNS may lose angular momentum, resulting in a slowly spinning BH with an accretion disk. The dynamics of magnetic disks around such slowly spinning collapsar BHs, along with their observational signatures, have so far remained unexplored.

We performed a series of 3D GRMHD simulations of collapsars to explore the origin of BH disk outflows as a function of BH spin, magnetization, and angular momentum of the progenitor star. We found that MADs around slowly spinning BHs generate magnetically dominated ($ \sigma \gtrsim 0.1 $) collimated outflows. For BH spins of $ a_{\rm crit} \lesssim 0.25 $, disk-driven outflows dominate over BZ jets, indicating that in most collapsar BH disks, the emission is primarily powered by the accretion disk rather than by the BH itself. Interestingly, this critical spin value is similar to that inferred for BHs that power lGRB jets. This suggests that even the most rapidly spinning BHs are likely to produce disk outflows that are comparably powerful to their BZ jets.

The disk outflows are driven by a comparable contribution of poloidal magnetic field lines threading the accretion disk (BP jets) and magnetic flux eruptions from regions of high magnetic pressure near the BH. Feedback from the cocoon, which is generated by the disk winds, induces a disk tilt that causes the outflows to wobble. With an isotropic-equivalent energy of $ E_{\rm iso} \approx 10^{52}\,\erg $, these outflows explode the star, leaving merely $ \sim 1\,\msun $ for BH mass growth. After breakout from the star, the disk outflows will generate multiple types of emission across the spectrum: shock breakout, cooling envelope, $ ^{56}{\rm Ni} $ decay, and synchrotron emission. We find that their cooling emission will be detectable out to hundreds of Mpc by ULTRASAT and $ D > {\rm Gpc} $ by UVEX and the Rubin Observatory. The detection rate of these events will provide valuable constraints on accretion disk physics and its formation rate in CCSNe.

The UV/optical cooling emission will be followed by $ ^{56}{\rm Ni} $ decay emission. Hydrodynamic simulations by \citet{Fujibayashi2023,Fujibayashi2024} suggest that the $ ^{56}{\rm Ni} $ emission is expected to last a few weeks, however, this result might change for magnetic disks\footnote{Our hydrodynamic disks fail to launch such powerful outflows (see \S\ref{sec:diverse}). This discrepancy may arise from the enforced axisymmetry in \citet{Fujibayashi2023,Fujibayashi2024}, which requires verification through 3D calculations.}. Additionally, recent neutrino--GRMHD simulations of collapsars indicate that strong magnetic fields in the accretion disk can facilitate the ejection of heavy $ r $-process elements from the disk \citep{Issa2024}. This could potentially drive fast ``kilonova'' emission, enhancing the signal we find, and altering the $^{56}{\rm Ni} $ yields compared to those found in hydrodynamic simulations. Future work incorporating neutrino transport in GRMHD simulations is necessary to investigate the impact of disk cooling on our results and model all three optical signals (cooling, kilonova, and $^{56}$Ni decay) from first principles.

The cooling and $ ^{56}{\rm Ni} $-powered peaks will jointly form a double-peaked optical emission \citep[e.g.,][]{Nakar2014}. UV/optical double-peaked light curves are common in SNe Ib/c \citep[e.g.,][]{Modjaz2009,Drout2016,Taddia2016,Ho2020,Gutierrez2021}, where the first peak appearing after several days at an absolute magnitude of $ M_{\rm AB} \approx -17 $ --- $ -18 $ and photospheric velocity $ \beta \lesssim 0.1 $, roughly consistent with our results, implying that disk winds may be its source. Future NUV and optical observations of SN early peaks will help distinguish between competing models for the fast early component -- whether driven by disk outflows \citep[see also][]{Hayakawa2018}, lGRB cocoon cooling emission \citep{Nakar2017}, long-lived jet--cocoon interaction \citep{Gottlieb2024c}, or other mechanisms.

Finally, recent GRMHD calculations suggest that high-density Rossby vortices in collapsar disks are promising sources of coherent gravitational waves \citep[GWs;][]{Gottlieb2024a}. Thus, our findings, which indicate powerful outflows producing strong electromagnetic emission, establish collapsar disks as promising multimessenger sources. In such systems, one messenger may enhance the detection prospects of the other, offering complementary insights into collapsar disk physics. Future work will examine the role of disk outflows in shaping Type Ib/c SN light curves, as well as the relationship between various electromagnetic emission components and the GW properties of collapsar disks.

\begin{acknowledgements}

We thank Amir Levinson and Aris Lalakos for useful comments.
J.B. acknowledges the support from the Simons Foundation Presidents Discretionary Fund. O.G. is supported by the Flatiron Research Fellowship. The Flatiron Institute is supported by the Simons Foundation. The computations in this work were, in part, run at facilities supported by the Scientific Computing Core at the Flatiron Institute, a division of the Simons Foundation.
This research used resources of the National Energy Research Scientific Computing Center, a DOE Office of Science User Facility supported by the Office of Science of the U.S. Department of Energy under Contract No. DE-AC02-05CH11231 using NERSC allocations m4603 (award NP-ERCAP0029085).
  
\end{acknowledgements}
		
	\bibliography{refs} 
	
\end{document}